\documentstyle[prb,aps,twocolumn,amssymb,amsfonts,epsfig]{revtex}

\begin{document}

%----------------------------------------------------------------------
% Title

\title{Spin-phonon interaction in doped high-T$_C$ superconductors from
density functional calculations.}

\author
{T. Jarlborg}

\address{D\'epartement de Physique de la Mati\`ere Condens\'ee,
Universit\'e de Gen\`eve, 24 Quai Ernest Ansermet, CH-1211 Gen\`eve 4,
Switzerland} 

\date{\today}
\maketitle

%----------------------------------------------------------------------
% Abstract
Effects of coupling between phonon distortions and stripe-like spin waves in the
CuO plane of HgBa$_2$CuO$_4$ are studied by band calculations. Local exchange enhancements
depend sensitively on the local structure around Cu sites. Interactions where spin waves
have twice as long wave length as phonon waves can lead to a 'dip' in the density of states 
(DOS) below the Fermi energy, $E_F$.
This type of interaction is compatible with several seemingly contradictory
observations among high-$T_C$ cuprates, like the isotope effect, anti-ferromagnetic
fluctuations, stripes and pseudogaps. It can also
account for a large $T_C$.
\begin{abstract}

\end{abstract}


%----------------------------------------
% Intro
Numerous studies of high-T$_C$ superconductors have revealed 
unusual  properties in the normal and superconducting states.
A common feature of the different Hg-, Bi-, Y- or La-based high-T$_C$ copper
oxides is the almost planar CuO sheets, which  
generates a cylindrical Fermi-surface (FS) in density functional (DF) band calculations.  
However, real undoped materials become anti-ferromagnetic (AF)
insulators and it is only within a rather narrow range of (usually hole-) doping
that high-T$_C$ superconductivity can be observed. This delicate balance of the
normal state property from being metallic or AF insulating leads naturally
to the AF pairing interactions \cite{pines} as an alternative to the conventional
electron-phonon coupling as a possible mechanism for high-T$_C$ superconductivity.
Complex isotope effects, d-wave gap symmetry, magnetic excitations, stripes and pseudo gaps all
contribute to the ''unconventional'' properties of these
materials \cite{hof,tsuei,mook,tran,norm,zheng,tim}. 
The concept of stripes has been developed from models of strong correlation \cite{zaan}, where 
Hubbard model calculations produce gaps in the
spectral functions \cite{fleck}. From DF calculations it
has been shown that ''stripe-like'' AF modulations (AFM) of spin-waves or phonons in the CuO-plane lead to gaps in the
one-particle DOS \cite{tjrap}.  Such modulations decrease the
total energy if the gap opens at $E_F$, which leads to
a connection between wave length and doping. Taken separately,
both waves (spin or phonon) give large coupling parameters (spin fluctuation or electron-phonon). 
This motivates a closer look at the possible interaction of such modes. 
We show that structural distortions modify local Stoner enhancement and hence the calculations confirm that
spin-phonon coupling is large. Secondly, it is shown that a double gap structure
in the DOS occurs when the wave lengths of the spin wave is twice that of the phonon mode.

Interactions between AFM fluctuations
and phonon modes are modeled for one of the structurally simplest high-T$_C$ materials,
HgBa$_2$CuO$_4$ (HBCO), where electronic structure calculations are done for ''frozen''
phonon and spin waves. 
The DF-calculations use the LMTO method based on the local spin
density approximation \cite{lda} (LDSA) for supercells of HBCO, including staggered magnetic fields to generate
spin-waves (''stripes'') or atomic displacements to model phonons. The calculations for the two types of
modes are similar, but atomic displacements generate multipolar changes of the potential and calculations
of changes of total energies require full potential methods. Long wave length spin modulations are simpler since
the change in potential is spherical within each atom. The calculations are still lengthy because
of the very large supercells. Therefore, in this work we refrain from calculating total energies, but we focus
on the exchange enhancements and double gap structures in the DOS. The details of the
methods can be found in refs. \cite{tjrap,san}.

A forced AF (on Cu) configuration in undoped HBCO, modeled by LSDA calculations, requires applied
field amplitudes of about 23 mRy to open a gap, cf. Table 1. If the Ba atoms are moved 0.1 a.u. 
 closer to the CuO plane, the critical field  is reduced to about 16, while 
the opposite movement requires about 30 mRy. Other distortions have similar effects, the Ba mode
is one example. When the atomic
distortions are modulated along the cell (but with no modulation of the applied AF field) one can notice that
the local AF Cu-moments are largest in a region where the Ba are closest to the CuO-planes. 
Another example is the ''half-breathing'' mode \cite{shen}, where 
 two planar oxygens are moved closer or further away from the Cu, so that Cu-sites along [1,0,0] alternatively
have two O-neighbours at +x and -x closer than normal (''compressed'' Cu), 
or more distant
than normal (''dilated '' Cu). The calculated exchange enhancement is roughly 50 \% larger on the dilated 
than on the compressed Cu.
The spin induced gap is situated at the stoichiometric band filling,
63 valence electrons per HBCO cell or at 1008 electrons for a 16 formula unit (f.u.) supercell, when the AF configuration
has no modulation. But as was
shown in ref. \cite{tjrap}, if there is a modulation (AFM or phonon) then the gap is always found at a band filling
corresponding to 2 electrons below the position of the gap in the unmodulated case. That is, at 1006 electrons
for the 16 f.u. cell,
at 502 instead of at 504 electrons for the half as long cell, etc.. 
 The gain in kinetic energy is largest if the 
 AFM wave has the largest moments coinciding with the optimal local distortions around Cu (Ba
close to the CuO plane, dilated Cu, etc.). The nodes of the AFM wave, where the moments are zero, 
coincide with the parts of the phonon which are most unfavorable to magnetism. Furthermore, this implies that  
 the AFM spin wave is twice as long as the wave of the phonon modulation. 
 
 As an example we consider a long
 supercell containing 16 f.u., oriented along [1,0,0], cf. fig 1. 
  According to the rules for gap positions
one expects a gap near 1006 electrons, caused by one complete spin modulation along this cell.
We call this gap a 'spin gap'. Simultaneously, there are two wave lengths of a phonon modulation within the
cell, which gives a gap (called 'phonon gap') at a band filling of 1004 electrons.
Such features are seen in the DOS from the band calculations. The gaps are complete with zero DOS in a narrow
energy interval, if the amplitudes of the atomic distortions or the magnetic fields are large. If the amplitudes
are moderate, the gaps are partial 'pseudo gaps', where the DOS is reduced to make 'dips' at the corresponding
band fillings.
 Unpolarized band calculations for different phonon modulations (Ba, apical-O,
''breathing-O'') show the expected pseudo gaps. When the cell contains both the double phonon wave and the AFM wave,
one should expect two gaps. The DOS shown in fig. 2, is calculated for a spin wave distribution along the cell which
generates a charge modulation of two wave lengths along the cell. The strongest gap is at $E_F$ (1006 electrons)
and it is caused by the spin wave. A second smaller 'dip' approximately 2 electrons below $E_F$ is caused by a component
$V_q$ from the charge distribution.
Note that no phonon distortion is needed to see the charge gap in this case, but for a paramagnetic calculation
with two wave lengths of a phonon, one can see a tendency for a gap near 1004 electrons, see fig. 2. A  similar
charge redistribution appears in this separate paramagnetic calculation as in the spin polarized case with only
the AFM wave.  
There seems to be a
mutual enforcement of spin and charge gap, as well as increased amplitude of the charge transfers
in calculations containing both the phonon and the AFM wave. 

Another example of this is the half-breathing modes within a 8 f.u. cell along [1,0,0]. Two half-breathing modes,
which are the shortest possible double phonon mode that can fit into this cell, induce similar charge transfers as 
the spin wave,
as seen in Table 2. The combination of
an unmodulated (AF) spin configuration {\it and} the phonon wave gives moments that are about 60 percent larger on the 
''dilated'' than on the ''compressed'' Cu, for the same applied field amplitudes on both types. 
 The ''compressed'' Cu atoms gain charge in the
paramagnetic calculations, just as the Cu at nonmagnetic node positions of a spin wave do. Both the phonon and the spin wave
 enforce the
$V_q$ coefficient for a (charge) gap (at 500 electrons for the 8 f.u. cell). 
This can interpreted as a type of valence fluctuation.
These have been suggested for a mechanism for exotic pairing and superconductivity \cite{bran}.
The results from the 8 f.u. cell show that they are coupled to the spin-phonon modes.

A few combinations of phonon and spin modulations
along [1,1,0] show partly similar trends.
There are 2 electrons per supercell separating the gaps induced by the spin- and phonon waves. For long wave lengths
(long supercells),
i.e. for underdoped cases, this means that the two gaps should appear relatively close to each other. 
 The energy difference between gaps
$\Delta E \approx n/N$, where $n$ and $N$ are the number of states and the DOS per f.u., respectively.
For instance, for the 16 f.u.
cell here, where n is 2/16 per f.u. and N is about 11 states/Ry per f.u. \cite{san}, 
the energy difference would be about 140 meV. For overdoped
cases, when the wave length is short, the two gaps should be more separated, although there is some compensation
from the increased DOS as $E_F$ approaches a van-Hove singularity at increased doping.

Photoemission \cite{hwu} and tunneling \cite{renn} measurements in the Bi-based material reveal a dip
at about 80 meV below $E_F$ even above $T_C$.
We suggest that this dip corresponds to the phonon gap of the DOS, as discussed above. The DOS at $E_F$ per CuO
plane in the Bi-compound is larger than in HBCO, and increasing further with doping \cite{san}.  The position
of the dip is of the correct order to correspond to the phonon dip. The origin of this dip
was much discussed a few years ago, because from tunneling it was shown to be pronounced only in the occupied part. This is
in qualitative agreement with the spin-phonon model. 

Not all phonon displacements produce double gaps. A mode with Cu displaced along $z$ makes
a relatively small difference between the moments on displaced and undistorted atoms. This distortion is symmetric
with respect to the CuO plane, and the wave length of an AFM modulation can be equal to
the wave length of the phonon. Spin and phonon gaps are weak and appear at the same energy.

 As was mentioned above, precise total energies are not calculated. 
However, we can make a number of qualitative observations of how a strong AF exchange enhancement
and a pseudo gap will change the total energy.
 The spin gap
is strongest if a system is close to an AFM transition
when there is almost no limit for the exchange enhancement. Vibrational amplitudes are of the order 0.1 a.u. 
at low T \cite{san}.  Normal
harmonic vibrations follow a parabolic potential, where the potential energy $U$ is a
quadratic function of the displacement $x$, $U = \frac{1}{2} K x^2$. $K$ is a force constant $d^2U/dx^2$. 
If, at low T and large $x$,
it is possible to gain energy because of a spin wave, it will also soften the phonon.
The quadratic behavior of $U$ will be flattened out ($K$ diminishes with $x$) which allows a further increase of $x$.
Mutual enhancements of distortion and moment will enforce both
gap structures. Several coupled spin-phonon modes may compete at large T,
when the free energy gain from a gap formation is less efficient because of the Fermi-Dirac smearing and thermal disorder.
A selection at lower T may leave one mode with the largest gain in energy to be
dominant with a more coherent effect on the DOS.
 
The Stoner enhancement $S$ is useful for a discussion of ferro magnetism. An extension for AF can be done via  
generalized spin susceptibility $\chi_q$. An approximate approach can be based on local $S$-factors
(defined within each site) calculated for AFM waves. Here they are 
 defined as the exchange splitting of the logarithmic derivative of the Cu-d band divided with
the applied field ($\epsilon$) at the same site.
The local Stoner factor $\bar{S}$ defines the local exchange integral $I$, $\bar{S} = NI = 1-1/S$.
An important decrease of the total energy is coming from the opening of a gap via the potential component $V_q$,  
which in the idealized case is close to
$-\frac{1}{2} N_q V_q^2 $, where $V_q$ is the band shift at $q$ and 
$N_q$ is a fraction of the total DOS that is affected by the splitting.
The total energy has also kinetic and exchange contributions from each site, which can be summed up,
if each site is independent of its neighbors.
\begin{equation}
\label{eq:etot}
E = \sum (N_i \epsilon^2 - N_i^2 I_i \epsilon^2) =  1/NS \sum m_i^2
\end{equation}
where $m=N\epsilon$ is the magnetic moment, and it is assumed that
$N$ and $S$ are not much varying from site to site.
The additional hybridization energy due to nonequivalent
neighboring atoms in the AF case is accounted for in local $S$-values, 
defined from spin-polarized results for the real spin-wave configuration. 
All these contributions make $S$ to vary non-linearly with the field, as can be seen in
Table 1 for the simple AF configuration of the undoped 2 f.u. cell. 
It is noted that the LSDA potential is too weak to explain 
a spontaneous AF order and a full gap. The local $I$ (or $N$) needs to be almost twice as large
to get a gap.
A similar reasoning can be made 
for the doped cases, with modulations of AFM waves.

Bardeen-Cooper-Schrieffer (BCS) theory without 
strong coupling effects is often sufficient to understand the mechanism for superconductivity based on
electron-phonon or paramagnon coupling  \cite{bcs,fay:80}.
A potential perturbation, as from a phonon, with Fourier component $V_q$ leads to a ''Cooper'' pairing
of electrons of opposite momentum $\pm q$. The BCS relation for $T_C$ contains the 
electron-phonon coupling $\lambda_{\rm ph}$;
\begin{equation}
\label{eq:tcep}
k_{B}T_{C} = 1.13 \hbar \Omega_D exp ({-1/ \lambda_{\rm ph}})
\end{equation}
where $\Omega_D$ is a characteristic phonon frequency. Here the coupling is for electronic states
of opposite spin. Since $\lambda_{ph} \sim NV^2/K = NV^2/M \Omega_D^2$, 
where $V$ is matrix element of the perturbation potential due to a displacement,
there is normally low coupling for stiff phonons. 
High phonon energies in combination
with fairly large $\lambda_{\rm ph}$, estimated from band theory, can hardly
explain the very large observed $T_C$'s in the copper oxides \cite{kra,san}, although
the role of phonons on superconductivity cannot be ignored \cite{hof,shen}.
An analogous formulation for equal spin pairing \cite{fay:80} is;
\begin{equation}
\label{eq:tcsp}
k_{B}T_{C} = 1.13 \hbar \omega_{sf} exp ({-1/ \lambda_{\rm sf}})
\end{equation}
The prefactor $\omega_{sf}$ is the frequency of the spin fluctuation and $\lambda_{\rm sf}$
is the coupling strength. 
These properties are in the ferro magnetic case related as \cite{mazin:97,jarlborg:86}:
\begin{equation}
\label{eq:somega}
\hbar \omega_{sf} = 2/NS~~~ ;~~~ \lambda_{\rm sf}=\frac{1}{2}S \bar{S}^2
\end{equation}

As for phonons there is an
antagonistic dependence of $T_{C}$ on frequency and coupling strength. For instance, if the spin coupling is large
via a large $S$-factor there is a softening, so that $\omega_{sf}$ and $T_c$ goes down.
The equal spin pairing has been suggested to give small $T_C$'s near
the FM transition in C15 compounds \cite{fay:80}.
Extreme softening leads to static spin for magnetic coupling or to a
structural transformation for phonon coupling.
The major contribution to the matrix element $V$ for an AFM perturbation comes from the opening of a gap.
An approximate value can be obtained from the gap model, where
the gap is equal to the maximum value of the exchange enhanced field. The model permits for an evaluation of
the matrix element $V \approx 1$, which with the denominator
 $K=d^2E/d\epsilon^2 = 2N/S$ makes
 $\lambda_{\rm sf} \approx \frac{1}{2}S$.
Eq. \ref{eq:etot} is formally as for ferro magnetic cases, whereby one can use $\hbar \omega_{sf} \approx 2/NS$.
Near the region for AF fluctuations, one can expect very
large $\lambda_{\rm sf}$, but unless a there is a mechanism for a large $\omega_{sf}$, it will not produce a large
$T_C$.

As was discussed above, 
Cu-moments can vary 50 percent or more along a phonon with realistic distortions. This implies strong
nonlinear variations as function of time and position, between low and high values of local Stoner enhancements 
and $\omega_{sf}$. If the development of a spin wave is strongly coupled to the atomic distortions,
as is suggested from the calculations, it is unlikely that the excitation spectrum for the spins would be
given by $\omega_{sf}$, since that quantity is independent of the phonons. It is more realistic if
the spin excitation spectrum is connected to
$\Omega_D$.
The nonlinear variations of $S$ permits a rapid development of a spin wave ($\omega_{sf} \rightarrow \Omega_D$) 
when the local $S$ is small,
alternated by strong coupling when $S$ is large.
Simultaneous large of values of the prefactor and $S$ can explain a much larger $T_C$'s than if 
$T_C$ is evaluated from spin fluctuations alone.  A major difficulty
for quantitative results from LSDA calculations is that the tendency towards AFM is too weak. As indicated above, the exchange
integral (or $N$) needs to be larger to explain a gap in the undoped system. If such an increase can be justified
for the modulated cases, it will make the difference between high and low $S$ values along a phonon distortion
even more spectacular. The average LSDA values for $\bar{S}$ are about 0.5 in the modulated cases, corresponding to a 
 $\lambda_{\rm sf}$ of the order 1. Omission of effects from strong coupling 
and non-linearity in a BCS-like formula with a frequency prefactor of 50 meV as for oxygen phonon modes, leads to a large $T_C$ of the
order 200 K. This is very approximate, but it shows qualitatively the possibilities of having a large $T_C$ from large AFM
enhancements in combination with high excitation frequencies. 
Impurities and strong electron phonon coupling are destructive to superconductivity based on the paramagnon mechanism 
\cite{fay:80}. However, with spin-phonon coupling there are
mutual enforcements between phonons and spin waves which can make $T_{C}$  more insensitive to perturbations
or disorder.  
Spin-phonon coupling would be 
consistent with a number of unusual observations among the high-$T_C$ materials such as an isotope
effect on T$_C$ and pseudo gaps, even in the case of pairing mediated by spin fluctuations. A crucial test of
spin-phonon coupling would be to measure the isotope effect on the spin excitation spectrum, since large
isotope shifts are expected from the model.

The calculations show that the conditions for AF depend sensitively on the local structure around Cu. This effect
is probably more important than in the present results, since LSDA calculations underestimate the tendency 
towards AF. Therefore, it can be concluded that spin-phonon coupling is important in these materials.
Similar charge transfers caused by phonons and spin waves, or charge transfers induced by stripe-like
spin waves alone, give a second gap in the DOS just below the main gap at $E_F$. The energy difference
between the gaps depends on doping, and as was shown previously \cite{tjrap}, the wave length of the stripes
are predicted to vary with doping as well.
Other effects from spin-phonon coupling are consistent with
 many observations in the high-$T_C$ oxides. In particular, this coupling might be important for the
mechanism of high-$T_C$ superconductivity.

%----------------------------------------
% Acknowledgments

%----------------------------------------------------------------------
% Biblio

%----------------------------------------------------------------------
% Table

%\end{document}
\begin{table}
\caption{Calculated local Stoner enhancement $S$ on Cu, gap in the DOS $E_g$ (mRy),
as function of the applied field (mRy) on Cu in undoped AFM Hg$_2$Ba$_4$Cu$_2$O$_8$
for normal structure and a distorted structure where Ba is moved 0.1 a.u.
towards the CuO plane.
}
% GS>
%\label{elphon}
\begin{tabular}{ccccc}
field (mRy) & normal & normal & dist. & dist.   \\
            & $S$ & $E_g$ &   $S$ & $E_g$ \\
\tableline
10 & 1.9 & - & 2.2 & - \\
15 & 1.8 & - & 2.1 & - \\
20 & 1.8 & - & 1.9 & 5 \\
25 & 1.7 & 3 & 1.8 & 12 \\
30 & 1.6 & 11 & 1.6 & 20  \\ 
\end{tabular}
\end{table}
%\end{document}
\begin{table}
\caption{Charges and moments on the two types of Cu in a 8 f.u. ''half-breathing'' modulation
along [1,0,0]. The O-displacement in the phonon is 0.1 a.u..
 Cu(1) are ''dilated'' and 
Cu(2) ''compressed'' sites, see the text. "AF" has no modulation of the anti-ferromagnetic
configuration. "AFM" is with modulation. 
}
% GS>
%\label{elphon}
\begin{tabular}{l c c c c c}\\
%\hline
&{case} &Charge &Moment &Charge &Moment \\
&{ } & Cu(1) & Cu(1) & Cu(2) & Cu(2) \\
\hline
&PM &10.32 &- &10.32 & - \\
&PM+phon. &10.26 &- &10.40 & - \\
&AF+phon. &10.25 &0.34 &10.40 &0.21 \\
&AFM+phon. &10.25 &0.34 &10.41 &0.0 \\
%\hline\\
\end{tabular}
\end{table}

%----------------------------------------------------------------------
% figure
\begin{figure}[tb!]
%\begin{center}

% figures directly (hr bw)\begin{figure}[tb!]
\leavevmode\begin{center}\epsfxsize8.6cm\epsfbox{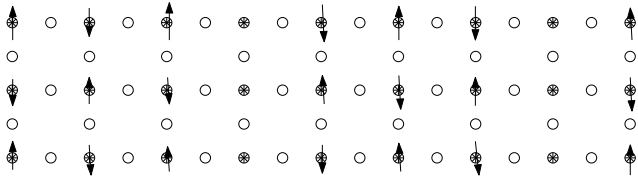}\end{center}
%\leavevmode\begin{center}\epsfxsize8.6cm\epsfbox{sa208x.ps}\end{center}
%\epsfxsize=120pt
%\centerline{\epsffile{sa208x.ps}}
%\epsfxsize=120pt
%\centerline{\epsffile{spcletfig2.eps}}
%\end{center}
\caption{
 The CuO plane for a supercell of 16 f.u. along [1,0,0] with modulated AFM moments.
The filled circles are Cu and open circles the plane oxygens.
The arrows indicate up- and down-moments on Cu.
 }
\end{figure}
\begin{figure}[tb!]

\leavevmode\begin{center}\epsfxsize8.6cm\epsfbox{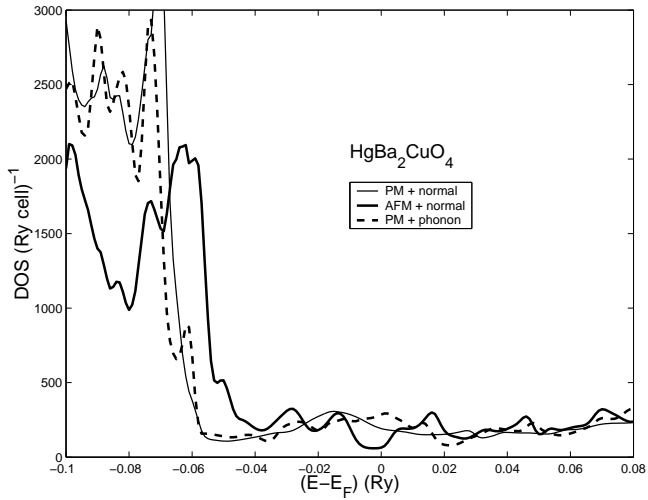}\end{center}
\caption{
 The total DOS for a 16 f.u. cell in fig. 1. The thin line is normal paramagnetic,
the heavy line
is with an AFM modulation, which makes a perturbation of the Fourier coefficient $V_q$ in the
spin potential.
$E_F$ is at 1006 electrons/cell (0.125 holes per f.u.).
 A second "dip" in the DOS is found at about 2 electrons/cell below $E_F$. The origin of this dip is found
 in a modulation of the charge (unpolarized Cu have more charge than Cu with moments) 
 which gives a potential perturbation in the coefficient $V_Q$, where
 $Q=2 \cdot q$. The broken line is for a paramagnetic calculation, where two wave lengths of ''half-breathing''
 distortions are introduced along the cell, see the text and table 2.
 }
\end{figure}

\end{document}